# Review and Analysis of Human Computer Interaction (HCI) Principles

**V. Hinze-Hoare**

Southampton University

## Abstract

The History of HCI is briefly reviewed together with three HCI models and structure including CSCW, CSCL and CSCR. It is shown that a number of authorities consider HCI to be a fragmented discipline with no agreed set of unifying design principles. An analysis of usability criteria based upon citation frequency of authors is performed in order to discover the eight most recognised HCI principles.

## Introduction

According to Diaper (2005) the chronology of HCI starts in 1959 with Shakel's paper on *"The ergonomics of a computer"* which was the first time that these issues were ever addressed. This was followed by Licklider who produced what has come to be known as the seminal paper (1960) on *"Man – Computer Symbiosis"* which sees man and computer living together. There was no further significant activity for almost 10 years when in 1969 the first HCI conference and first specialist journal, *"The International Journal of Man-Machine Studies"* was launched. The 1980s saw the launch of three more HCI journals and conferences with an average attendance of 500 (Diaper 2005). It was not until the 1990s that the "I" in HCI switched from *"interface to "interaction"* reflecting the vastly expanding range of digital technologies. It was also during the 1990s that the term "Usability" has come to be synonymous with virtually all activities in HCI. Prior to this HCI encompassed five goals to develop or improve:

- Safety
- Utility
- Effectiveness
- Efficiency
- Usability

Originally usability was the least but has since been promoted to cover everything. *"The study of HCI became the study of Usability"* (Diaper, 2005).

Brad Myers (1998) has reviewed the history of HCI from a technological point of view and shows that HCI started with university research in direct manipulation of graphical objects as long ago as 1960, with commercial research not starting until 1970 and commercial products available from 1980. Myers also highlights up and coming areas of modern HCI research

- **Gesture Recognition:**
  pen-based input device,
- **Multi-Media:**
  multiple windows and integrated text and graphics
- **3-D:**
  ultrasonic 3D location sensing system
- **Virtual Reality and "Augmented Reality":**
  much of the early research on head-mounted displays and on the DataGlove was supported by NASA.
- **Computer Supported Cooperative Work.**
  the remote participation of multiple people at various sites
- **Natural language and speech:**
  fundamental research for speech and natural language understanding and generation

## The Basic Characteristics and Structure of HCI

Dix *et al* (1992) states that "*Human computer interaction can be defined as the discipline concerned with the design, evaluation, and implementation of interactive computing systems for human use and with the study of major phenomena surrounding them"*

HCI has become an umbrella term for a number of disciplines including theories of education, psychology, collaboration as well as efficiency and ergonomics as shown in Figure 1.

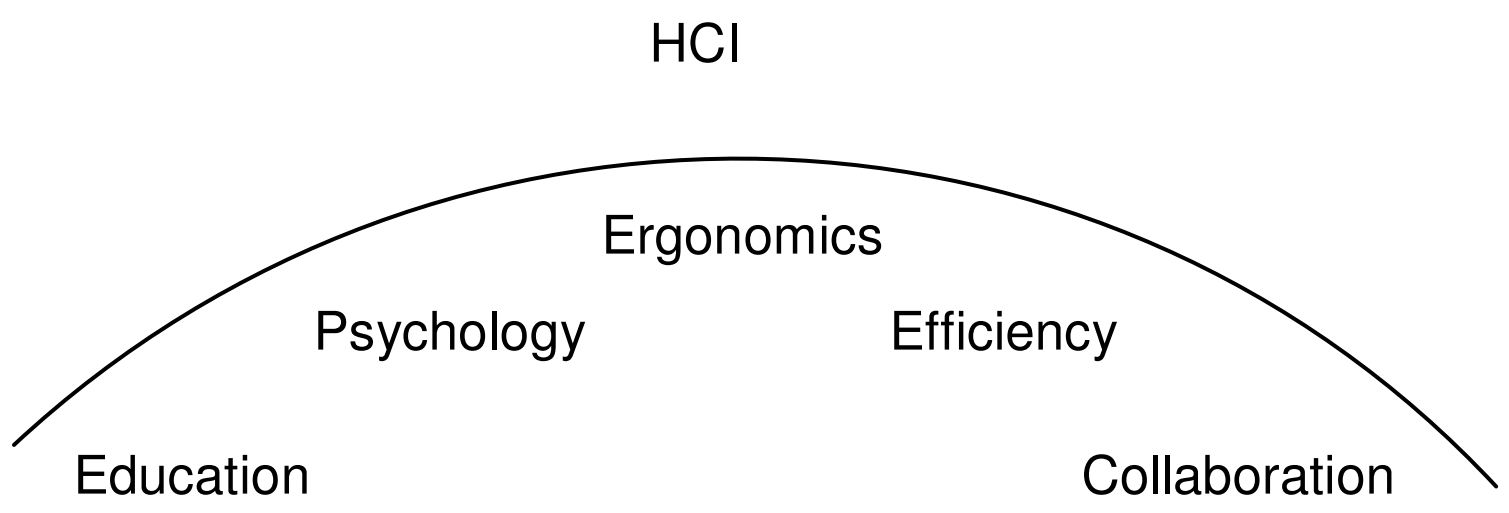

**Figure 1 HCI Components**

Recent developments in the area of HCI have shown an interest in adaptive interfaces (Savidis and Staphanidis 2004), speech recognition (Wald 2005), gestures (Karam and Schraefel 2005) and the role of time (Wild and Johnson 2004, and Oulasvirta and Tamminen 2004).

Recently Hinze-Hoare (2006) has contended that HCI encompasses the sub domains of Computer Supported Collaborative Working CSCW, Computer Supported Collaborative Learning CSCL, and Computer Supported Collaborative Research CSCR where each domain is a subset of the previous one. See Figure 2.

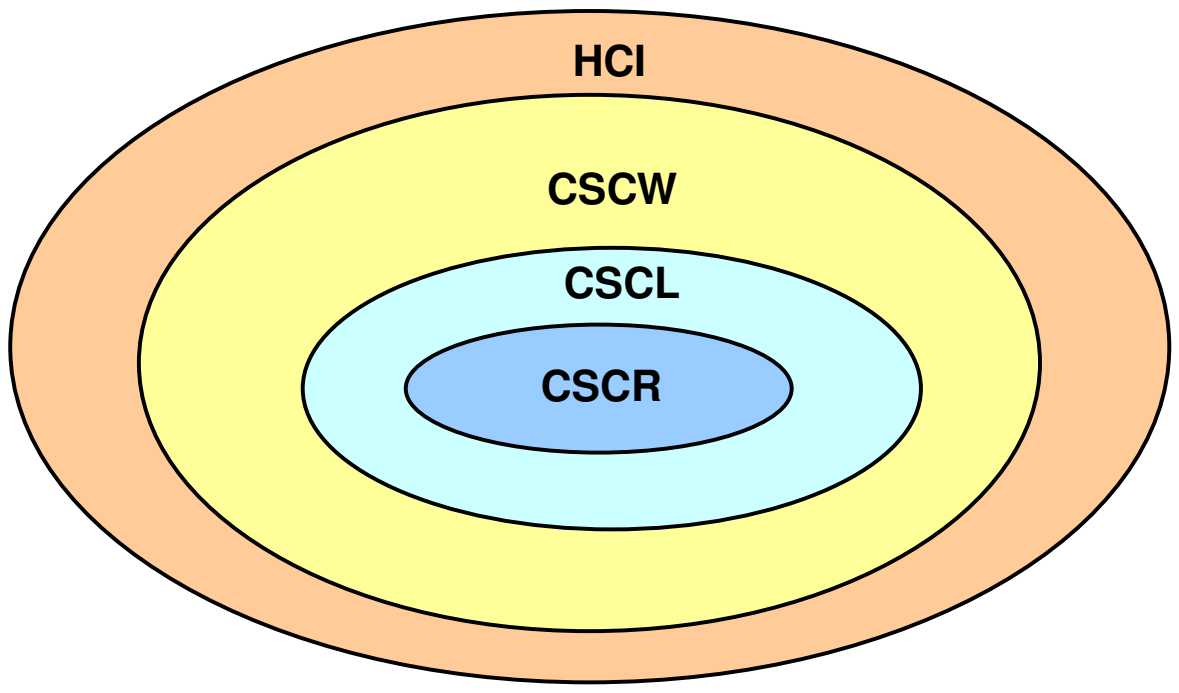

**Figure 2 HCI Domain Relationships**

## HCI Theories and Principles

There are typically many thousands of rules which have been developed for the assessment of usability (Nielsen, J. 1993, p19), and there have been many attempts to reduce the complexity to a manageable set of rules (Nielsen, J. 1993, Baker, Greenberg and Gutwin, 2002. Jacob Nielsen has produced 10 rules which he calls usability heuristics and which are designed to explain a large proportion of problems observed in interface design, which he recommends should be followed by all user interface designers.

**1. Simple and natural dialogue**
Efforts should be made to avoid irrelevant information. Nielsen says that every extra unit of information competes with units of relevant information and diminishes its visibility.

**2. Speak the Users' language**
All information should be expressed in concepts which are familiar to the user rather than familiar to the operator or the system.

**3. Minimize the Users' memory load**
It is important that the user should not have to remember information from one part of a dialogue to another. Help should be available at easily retrievable points in the system.

**4. Consistency**
Words situations and actions should always mean the same thing no matter where they occur in the system.

**5. Feedback**
Users should always be informed about what is going on in the system in a timely and relevant way.

**6. Clearly marked Exits**
Errors are often made in choosing functions which are not required and there needs to be a quick emergency exit to return to the previous state without having to engage in extended dialogue.

**7. Shortcuts**
Required by the expert user (and unseen by the novice user) to speed the interaction with the system.

**8. Good error messages**
These need to be expressed in a plain language that the user understands which are specific enough to identify the problem and suggest a solution.

**9. Prevent Errors**
A careful design will prevent a problem from occurring.

**10. Help and documentation**
the best systems can be used without documentation. However, when such help is needed it should be easily to hand, focused on the users task and list specific steps to solutions.

Baker, Greenberg and Gutwin (2002) have taken Jakob Nielsen's heuristic evaluation a stage further and considered the problems posed by groupware usability concerns. They have adapted Nielsen's heuristic evaluation methodology to collaborative work within small scale interactions between group members. They have produced what they call 8 groupware heuristics.

- **Provide the means for intentional and appropriate verbal communication.**
  The most basic form of communication in groups is verbal conversation. Intentional communication is used to establish common understanding of the task at hand and this occurs in one of three ways.
    - People talk explicitly about what they are doing
    - People overhear others conversations
    - People listen to running commentary that people produce describing their actions.
- **Provide the means for intentional and appropriate gestured communication.**
  Explicit gestures are use alongside verbal communication to convey information. Intentional gestures take various forms. Illustration is acted out speech, Emblems are actions that replace words and Deixis is a combination of gestures and voice

- **Provide consequential communication of an individual's embodiment**
  Bodily actions unintentionally give off information about who is in the workspace, where they are and what they are doing. Unintentional body language is fundamental for sustaining teamwork.
- **Provide consequential communication of share artefacts**
  A person manipulating an artefact in a workspace unintentionally gives information about how it is to be used and what is being done with it
- **Provide Protection**
  People should be protected from inadvertently interfering with the work of others or altering or destroying work that others have done
- **Manage the transitions between tightly and loosely coupled collaboration**
  Coupling is the degree to which people are working together. People continually shift back and forth between loosely and tightly coupled collaboration as they move between individual and group work
- **Support people with the coordination of their actions**
  Members of a group mediate their interactions by taking turns negotiating the sharing of a common workspace. Groups regularly reorganize the division of work based upon what other participants have done or are doing.
- **Facilitate finding collaborators and establishing contact**
  Meetings are normally facilitated by physical proximity and can be scheduled, spontaneous or initiated. The lack of physical proximity in virtual environments requires other mechanisms to compensate.

Others have produced alternative sets of rules. However the important issue is that there is no consensus as to which set of rules should be applied in any given case. In other words HCI is a fragmented discipline which according to Diaper (2005) shows a lack of coherent development.

# HCI Models

A variety of different models have been put forward which are designed to provide an HCI theory in a particular context. This includes Norman's Model, Abowd and Beale's model and the audience participation model of Nemirovsky (2003) which presents a new theoretical basis for audience participation in HCI.

## Norman's model of interaction

This has probably been the most influential (Dix *et al* 1992 p105) because it mirrors human intuition. In essence this model is based on the user formulating a plan of action and then carrying it out at the interface. Norman has divided this into seven stages:

1. establishing the goal
2. forming the intention
3. specifying the action sequence
4. executing the action
5. perceiving the system state
6. interpreting the system state
7. evaluating the system state with respect to the goals and intentions

## The Interaction Model

Abowd and Beale (Dix *et al* 1992 p106) have produced an interaction framework built on Norman's model but theirs is designed to be a more realistic model.

This has four main components:

1. the system S
2. the user U
3. the input I
4. the output O

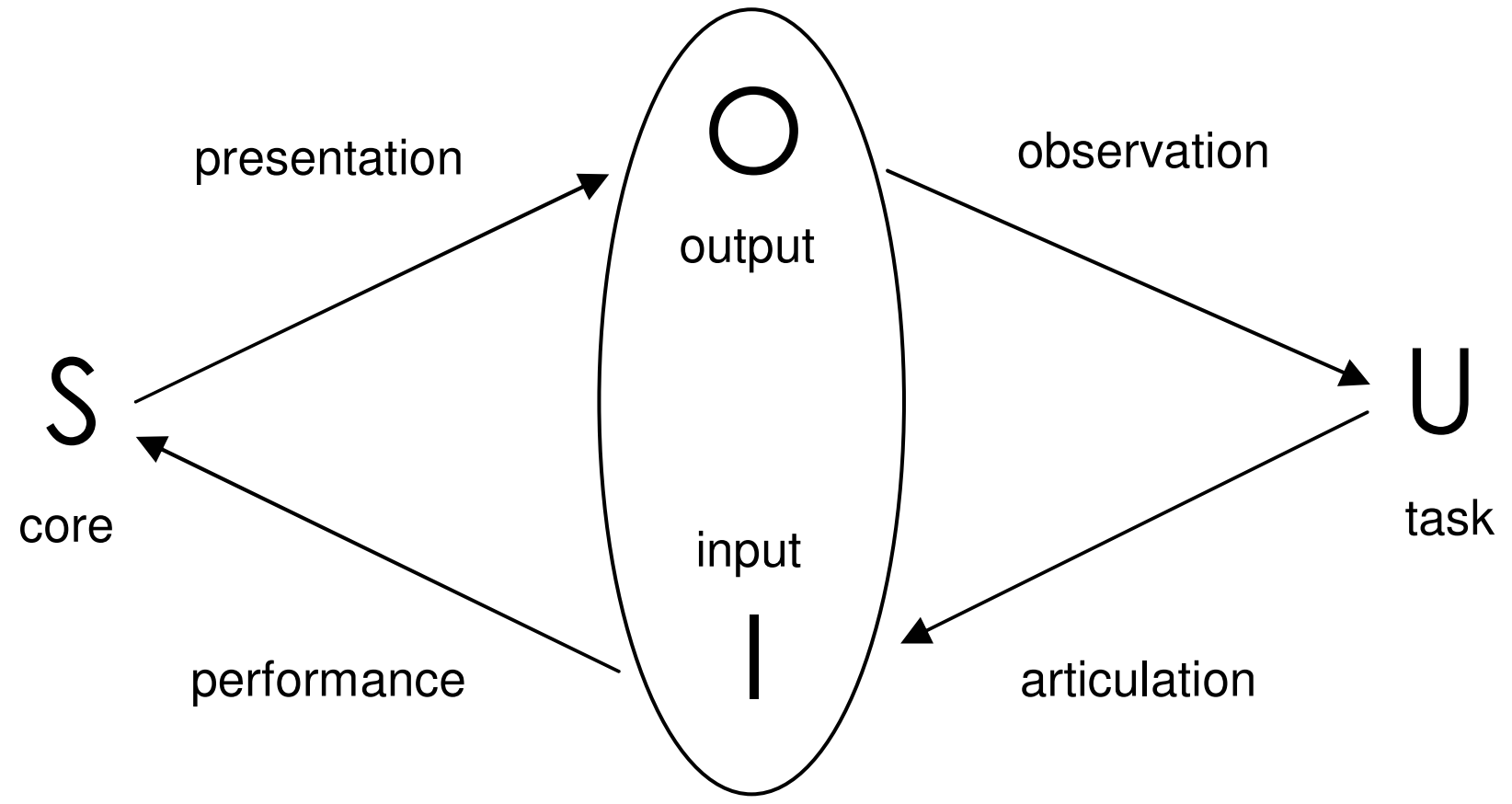

**Figure 2 The General Interaction Framework**

The interface sits between the user and the system and there is a four step interactive cycle as shown in the labelled arrows of **Error! Reference source not found.** The user formulates a task to achieve a goal. The user manipulates the machine through the input articulated by the input language. The input language is translated into the systems core language to perform the operation. The system is in a new state which is featured as the output. The output is communicated to the user by observation.

## Audience Participation Model

Nemirovsky (2003) considers that the old perspective is that of computers as deterministic boxes blindly following their commands while users are incapable of changing the course of the program running on the computer. To this he presents an alternative and proposes that users should be considered as an audience rather than participants. Old models include the idea that the mass of people wish to be entertained rather than to be creative and are punished for any creative thinking while using a computer which is regarded as making a mistake. As a consequence computer users do not have a proper framework to express themselves. This is a strikingly radical approach. Instead Nemirovsky is concerned with users as an audience that explore the media space. He goes on to discuss the emonic environment which he defines as a framework for creation, modification, exchange and performance of audio visual media. This is composed of the three layers:

- Input (interfaces for sampling)
- Structural (a neural network for providing structural control)
- Perceptual (direct media modification)

This latter and more radical model is unlikely to have any real application to the CSCL problem at hand, since its sphere of application is designed to emulate the creative inspiration of the artist battling against the mechanical controls of the machine.

## HCI Analysis Methodology

A number of different methodologies have been created to determine the effectiveness of HCI measurements.  These have been refined resulting in the *User Needs Analysis* of Lindgaard *et al* (2006) that suggests how and where user centred design and requirements engineering approaches should be integrated. After reviewing various process models for user centred design analysis they suggest a refined approach and identified the main problems as:

- The decision where to begin and end the analysis needs to be clarified.
- Deciding how to document and present the outcome

Lindgaard's user needs analysis method involves the following steps

- **First**: Identify user groups and interview key players from all groups to find the different roles and tasks of the primary and secondary users
- **Second**: Communicate this information to the rest of the team by constructing task analysis data and translating this into workflow diagrams supporting the user interface design. Create a table that shows the information about user roles and data input
- **Third**: Upon submitting the first draft of the user needs analysis report create the first iterative design prototype of the user interface based on minimising the path of data flow. (Initially prototyping in PowerPoint was faster and more effective that prototyping in Dreamweaver).
- **Fourth**: Prototypes were handed over to developers as part of the user interface specification package.
- **Fifth**: Usability testing was used to determine the adequacy of the interface. Feedback from watching users work with the prototype and discussing with them what they were doing always resulted in more information.
- **Sixth**: Prototype usability testing meant that the requirements became clearer which resulted in more changes to the user interface design and the prototype.
- **Seventh**: The formal plan involved three iterations of design- prototype- usability test for each user role (they could not keep to this and had no more that two test iterations and in most cases only one)
- **Eighth**: Practical issues of feasibility should not be overlooked in the quest to meet users' needs. A highly experienced software developer is a necessity on the user interface design team in order to ensure that the changes were proposed were feasible ( in some cases interface ideas were dropped because they were not feasible, take too long or cost too much).

# The Fragmentation of HCI

The History of HCI according to Diaper (2005) shows a lack of coherent development. There is no agreement as to

- What HCI should be
- What HCI can do
- How HCI can do it
- How HCI can be allowed to do it

The fragmentation of HCI discipline is already so extensive according to Diaper 2005 that it is hard to even characterise the method of approach. As an example different practitioners have different priorities and different methodologies. Some approaches will start with requirements, while others start with evaluation and yet others with dialogue or user modelling or scenarios or information or design or artefacts or processes. This lack of agreement highlights the necessity for the development of a general systems model, both in the general HCI approach and in the specific collaborative approach. This view is also expressed by a number of other researchers:

**Kligyte**, **(2001)** has recorded that "*CSCL emerged as an autonomous research field out of the wider CSCW research area quite recently, and there is still lack of consensus about core concepts, methodologies and even the object of study".*

**Lipponen (2002)** concludes his review paper on CSCL with the comment "*There is still no unifying and established theoretical framework, no agreed objects of study, no methodological consensus or agreement about the concept of collaboration or unit of analysis".*

**Strijbos *et al* (2005)** have pointed out that a review of CSCL conference proceedings revealed a general vagueness in the definitions of units used in the analysis of CSCL. They further comment that arguments were lacking in choosing units of analysis and reasons for developing content analysis procedures were not made explicit. They conclude that CSCL is still an emerging paradigm in educational research. This suggests that there may be a need to evaluate new definitions in order to contextualise this work.

# The Analysis of HCI principles

It has been shown that HCI is in a state of fragmentation. This leads to the problem of adopting a coherent and consistent set of principles by which to measure the HCI performance of an interface. To this effect many sets of principles have been put forward by many different authorities in this field. However there is no consistent single set of principles accepted by all. The purpose of this section is to normalise the range of principles which have been proposed and to determine the most significant set.

## Methodology

It will be described here how the normalisation process is accomplished.

**Stage1: Use Authors' citation frequency as a weighting factor.**
A survey of the HCI literature was undertaken based upon the citation frequency of authorship. It was deemed that the most frequently cited authors would provide the most important and respected HCI principles. Furthermore the citation frequency of the authors would constitute a mechanism for weighting the authority of the principles. HCIBIB.ORG maintains a citation frequency database of all HCI authors[1]. This shows the most frequently cited authors (10 or more publications) in the HCI Bibliography starting in Dec 1998 generated from the author fields and from tables of contents. The top ten names together with their citation frequency are given as follows[2]:

| Citation Number | Author | Weighting |
|---|---|---|
| 436 | Nielsen, | 29. |
| 186 | Shneiderman, | 13 |
| 165 | Carroll, | 11 |
| 133 | Myers, | 9 |
| 102 | Salvendy, | 7 |
| 96 | Pemberton, | 7 |
| 92 | Marcus, | 6 |
| 92 | Grudin, | 6 |
| 87 | Perlman, | 6 |
| 87 | Greenberg, | 6 |

**Table 1 HCI Authors and their Citation Ranking**

These are simply weighted as a percentage of the overall number of citations so that in effect Nielsen is showing 29% of the total number of citation listed in Ranking. The work of each significant author was examined for HCI principles and these principles listed in a matrix and factored according to HCI categories. By this means a full set of HCI principles was drawn from the works of each significant author. The most popular of these principles were obtained from many different authors while some of the least popular principles were drawn from just one or two authors. Every principle had at least one author proposing it.

[1] http://www.hcibib.org/authors.html
[2] As of March 2007

**Stage 3: Determine the weighted frequency of HCI principles**

The number of times that a particular HCI principle was proposed by a significant author multiplied by a weighting factor derived from the author citation frequency allowed a ranking of HCI principles to be determined.

It is expected that this method will overcome some of the degrees of fragmentation of the HCI field by bringing together a set of principles which have been constructed in such a way as to reflect the degree of respect and authority attributable to the authors who proposed them.

This analysis provides what is thought to be first general approach at consolidating HCI principles in this way see Table 2.

| Weighting | Dix et al-Principles p162 | Schneiderman Principles – Desinging the User Interface | Jenny Preece- Principles | Jenny Preece- Interaction Design quoting from Medline Plus | Donald Norman - The Design of Everyday Things | Robert J. Kamper-Extending Usability of Heuristics (Journal of HCI) | Smith & Mosier (1986), Schneiderman P 80 | Lockhead Principles (1981) Schneiderman P 80 | Caroll, Hollan et al (2000) ACM Transaction,Vol 7, No.2, June 2000 | Ken Maxwell, Carrol P 191, Levels of HCI maturity | Steve Pemberton http://www.cwi.nl/~steven/ | Brad A. Myers | Jeff Raskin, (2000) The Human Interface | Jennifer Niederst, 1999, Web Design in a Nutshell P 23/4 | Cogdill, K * 1999, Medline plus Interface evaluation, Final Report, University of Maryland | Jacob Nielsen 2001 Ten Usability heuristics | Mill & Shultz, Carroll P 537 - The next Frontier for HCI research | Andrew Dillon, Carroll P 466 Do we really know our users? | Erikson & Kellog, Caroll P 326 | Raw | weighted |
|---|---|---|---|---|---|---|---|---|---|---|---|---|---|---|---|---|---|---|---|---|---|
| Weighting | 1 | 13 | 1 | 1 | 1 | 1 | 1 | 1 | 11 | 1 | 7 | 9 | 1 | 1 | 1 | 29 | 1 | 1 | 1 | Raw | **weighted** |
| Predictability | | | | | 1 | 1 | | | | | | | | | | 1 | | | 1 | 4 | **32** |
| Synthesisability | | | 1 | | | 1 | 1 | 1 | | 1 | | | | | | 1 | | | | 6 | **34** |
| Familiarity | | 1 | 1 | 1 | 1 | 1 | 1 | 1 | | | 1 | | | | 1 | 1 | 1 | | | 11 | **57** |
| Generalisability | | | 1 | | | | | | | | | | | | | | | | | 1 | **1** |
| Consistency | | 1 | 1 | 1 | 1 | 1 | 1 | 1 | | | 1 | | | | 1 | 1 | 1 | | | 11 | **57** |
| Feedback | | | | 1 | | 1 | | | | | | | | | 1 | | | | | 3 | **3** |
| | | | | | | | | | | | | | | | | | | | | | |
| Dialogue initiative | | 1 | 1 | 1 | 1 | | 1 | 1 | | | | | | | 1 | | | | | 7 | **19** |
| Multithreading | | | | | | | | | | | | | | | | | | | | 0 | **0** |
| Task Migrateability | | | 1 | | | 1 | | | | 1 | 1 | | | | | 1 | 1 | | | 6 | **40** |
| Substitutivity | | 1 | 1 | 1 | | 1 | | 1 | | | 1 | | | | 1 | 1 | | | | 8 | **54** |
| Customizability | | | 1 | | | 1 | 1 | | | | 1 | | | | | | 1 | | | 5 | **11** |
| | | | | | | | | | | | | | | | | | | | | | |
| Observability | | 1 | | | 1 | 2 | | | | | | | | | | | | | | 4 | **16** |
| Recoverability | | 2 | 2 | 2 | 1 | 2 | | | | 2 | | | 1 | | 2 | 2 | | | | 16 | **96** |
| Responsiveness | | 1 | 1 | | | | | | | | | | 1 | | | | 1 | | 1 | 5 | **17** |
| Task Conformance | | | | | | | | | | 1 | | | 1 | | | | | | | 2 | **2** |
| | | | | | | | | | | | | | | | | | | | | | |
| Social Ergonomics | | | | | | | | | 1 | | | | | | | | | 1 | 1 | 3 | **13** |
| Cultural Ergonomics | | | | | | | | | 1 | | | | | | | | | | | 1 | **11** |
| Holistic Ergonomics | | | | | | | | | | 1 | | | | | | | | | | 1 | **1** |
| Physical Ergonomics | | | | | 1 | 1 | | | | | | | | | | | | 1 | | 3 | **3** |
| Perceptual Ergonomics | | | | | | 1 | | | | | | | | | | 1 | | 1 | | 3 | **31** |
| Cognitive Ergonomics | | | | | | | | | 1 | | | | | | | | | 1 | | 2 | **12** |
| | | | | | | | | | | | | | | | | | | | | | |
| Economic Accessibility | | | | | | | | | | | | | | 1 | | | | | | 1 | **1** |
| Technical Accessibility | | | | | | | | | | | | | | | | 1 | | | | 1 | **29** |
| Visual Disability | | | | | | | | 1 | | | | | | 1 | | | | | | 2 | **2** |
| Auditory disability | | | | | | | | | | | | | | | | | | | | 0 | **0** |
| Speech disability | | | | | | | | | | | | | | | | | | | | 0 | **0** |
| Motor disability | | | | | | | | 1 | | | | | | 1 | | | | | | 2 | **2** |
| Cognitive disability | | | | | | | | 1 | | | | | | | | | | | | 1 | **1** |

**Table 2: Frequency analysis of HCI principles**

## Findings

The fundamental principles of each author were examined, categorised and weighted according to the citation frequency and the top eight rules were found to be:

| | | |
|---|---|---|
| 1 | Recoverability | 96 |
| 2 | Familiarity | 57 |
| 3 | Consistency | 57 |
| 4 | Substitutivity | 54 |
| 5 | Task Migratablility | 40 |
| 6 | Synthesisability | 34 |
| 7 | Predictability | 32 |
| 8 | Perceptual Ergonomics | 31 |

**Table 3: Weighted HCI rules according to frequency of use**

In detail these eight principles are as follows:

**1. Recoverability**

This is the ability of users recovering from their errors, which they invariably make. There are two directions in which recovery can occur both forward and backward. Forward error recovery involves the prevention of errors. Backward error recovery concerns the easy reversal of erroneous actions. The latter is usually concerned with the user's actions and is initiated by the user. The former is one, which should be engineered into the system and initiated by the system. In this sense recoverability is connected to fault tolerance, reliability and dependability. Ken Maxwell (2001) considers this basic usability a level one priority, which he calls error protection. Jeff Raskin (2000) rates this as part of his first law of interface design, which states, "a computer shall not harm your work or through inaction allow your work to come to harm". This gained a weighted rating of 96.

**2. Familiarity**

This is the degree to which the user's own real world personal experience and knowledge can be drawn upon to provide an insight into the workings of the new system. The familiarity of a user is a measure of the correlation between their existing knowledge and the knowledge required to operate the new system. To a large extend familiarity has its first impact with the users' initial impression of the system and the way it is first perceived and whether the user can therefore determine operational methods from his own prior experience. If this is possible this greatly cuts down the learning time and the amount of new knowledge that needs to be gained. The term familiarity is proposed by Dix *et al* (1992) but is referred to by other authors under different terms i.e. as guessability. Schneiderman (1998) and Preece (1994) each refer to familiarity in terms of the reduction of cognitive load. This was the most quoted principle amongst all HCI authors and as such gained a weighted rating of 57.

**3. Consistency**

Consistency, according to Dix *et al* (1992) relates to the likeness in behaviour arising from similar situations or similar task objectives. He also thinks that this is probably the most widely mentioned principle in the literature on user interface design. This principle comes out as joint first place with familiarity. It is considered of vital importance that the user has a consistent interface. However, there is an intrinsic difficulty in defining the nature of consistency, which can take many forms. Consistency is relative to a particular area for example one can speak of consistency of mouse movements, menu structures, response etc. Whereas familiarity can be considered as “consistency with respect to personal experience” this consistency is one with respect to “internal similarity of appearance and behaviour”. This principle shared the top slot with familiarity, also with a weighted rating of 57.

**4. Substitutivity**

This concerns the ability of the user to enter the same value, or perform the same action in different ways according to the user’s own personal preference. For example a user might wish to enter values in either inches or centimetres, or he may wish to open a program with the mouse or with the keyboard. This input Substitutivity contributes towards an overall flexible HCI structure, which allows the user to choose whichever he considers most suitable. Schneiderman (1998) and Preece (1994) provide a specific example of providing shortcuts as an alternative. This is the ability of the interface to provide multiple methods for performing the same task. This achieved a weighted rating of 54.

**5. Task Migratablility**

This concerns the transfer of control for executing tasks between the system and the user. Checking the spelling of a document is a good example. The user can quite easily check the spelling for himself by the use of a dictionary. However the task is made considerably easier if it can be passed over to the system to perform with simple checks made by the user as to the acceptable spelling i.e. the difference between US and British Dictionaries. This is an ideal task for automation. However, it is not desirable to leave it entirely in the hands of the computer as dictionaries are limited and therefore the task needs to be handed over to the user at those complex points where the system cannot cope. Ken Maxwell (2001) talks of this as level two collaborative organisational interaction which he considers being a high level of HCI interaction.

This is the ability of the interface to hand the task over to the user so that the initiative rests with the human side of the interaction. This can be measured by the degree of performance available through the use of unfamiliar tasks. This has a weighted rating of 40.

**6. Synthesisability**

This is the ability of the interface to allow the user to construct a predictive mental model of how it operates. In other words through using the interface the user gains an understanding of what to expect next (predictability). In addition the user works out a framework or scaffolding for all the actions he can perform. For example if the user moves a file from one place to another he should be able to check after the action is completed that the file is in the new location as expected. This is what Dix *et al* (1992) call the “honesty of the system”. Without this the user would not be able to learn the consistent procedure for interacting with the interface. This has a weighted rating of 34.

**7. Predictability**

This is support for the user to determine the effect of future actions based upon a past knowledge of the system. It allows the user to know beforehand what will happen when he clicks on a menu item or presses a key. This is a user centred concept where the user can take advantage of his knowledge of how the system is going to respond. Any system which does not respond as expected or responds inconsistently will be difficult if not impossible to learn. This has a weighted rating of 32.

**8. Perceptual Ergonomics**

Human perception involves the stimulus of sense organs. Measuring the Ergonomic properties of stimulus patterns is one method by which a more efficient interface can be created. This places the emphasis on the human side of HCI. For instance if human hearing cannot perceive very high notes then it would be important to ensure that audible signals did not fall outside the human range. Similarly if the user cannot perceive particular colours then those colours must be removed from the interface. Tracking the way humans perceive things is important to making an interface efficient for human use. This had the lowest weighted rating of 31.

## HCI summary

It has been shown that HCI theories are not yet fully established and that the discipline is highly fragmented making it difficult to characterise a single method of approach or even a set of accepted principles. The lack of agreement between authorities in this field suggests that the approach must be carefully tailored to the specific needs of the environment to which they are applicable.

This section has briefly considered the history of HCI which showed how usability has become the central feature of virtually all HCI activities from the 1960's onwards. The structure of HCI has been reviewed to show how it encompasses a number of disciplines. Three HCI models were examined which illustrated the increasing refinement of interactive description culminating in Abowd and Beale's interaction theory. The approach to HCI analysis has evolved into the methodology of Lindgaard *et al* which focuses on user's needs. This is an approach which is commonly adopted and it will be addressed in more detail when the methodology of this project is considered.

Because of the fragmentation of HCI principles it was felt necessary at this stage to perform a frequency analysis of HCI authors and their chosen principles. This was done on the basis of the key features that each author listed as being the most important. These were then weighted according to the citation frequency of the authors themselves. The purpose of this was to produce a set of principles which would be held to be the most accepted. It was found that eight rules have been established by this analysis.